# A Single Visualization Technique for Displaying Multiple Metabolite-Phenotype Associations

Mir Henglin[1*], Teemu Niiranen[2*], Jeramie D. Watrous[3], Kim A. Lehmann[3], Joseph Antonelli[1.4], Brian L. Claggett,[1] Emmanuella J. Demosthenes,[1] Beatrice von Jeinsen[2], Olga Demler,[5] Ramachandran S. Vasan[2,6], Martin G. Larson[2,7], Mohit Jain[3*], Susan Cheng[1,2*]

*Denotes equal author contribution

From the [1]Cardiovascular Division, Department of Medicine, Brigham and Women's Hospital, Harvard Medical School, Boston, MA; [2]Framingham Heart Study, Framingham, MA; [3]Departments of Medicine & Pharmacology, University of California San Diego, La Jolla, CA; [4]Biostatistics Department, Harvard School of Public Health, Harvard University, Boston, MA [5]Preventive Medicine, Department of Medicine, Boston University Medical Center, Boston, MA; [6]Preventive Medicine, Department of Medicine, Brigham and Women's Hospital, Boston, MA; and, [7]Biostatistics Department, School of Public Health, Boston University, Boston, MA

**Correspondence:** Susan Cheng, MD, MPH, Brigham and Women's Hospital, Harvard Medical School, scheng@rics.bwh.harvard.edu; and, Mohit Jain, MD, PhD, University of California, San Diego, mjain@ucsd.edu.

**Running Title:** Rain plot visualization technique

**ABSTRACT**

More advanced visualization tools are needed to assist with the analyses and interpretation of human metabolomics data, which are rapidly increasing in quantity and complexity. Using a dataset of several hundred bioactive lipid metabolites profiled in a cohort of ~1500 individuals sampled from a population-based community study, we performed a comprehensive set of association analyses relating all metabolites with eight demographic and cardiometabolic traits and outcomes. We then compared existing graphical approaches with an adapted 'rain plot' approach to display the results of these analyses. The rain plot combines the features of a raindrop plot and a parallel heatmap approach to succinctly convey, in a single visualization, the results of relating complex metabolomics data with multiple phenotypes. This approach complements existing tools, particularly by facilitating comparisons between individual metabolites and across a range of pre-specified clinical outcomes. We anticipate that this single visualization technique may be further extended and applied to alternate study designs using different types of molecular phenotyping data.

## INTRODUCTION

Advancements in metabolomics technologies have enabled the generation of large-scale metabolomics measures in human studies.[1,2] Accordingly, newer generation visualization tools are needed to assist with the analyses and interpretation of these increasingly high-dimensional and complex data sets. Several resources now offer a variety of techniques for visualizing metabolomics data structure and exploring the inter-relations between individual and groups of metabolites.[3-16] The most commonly used methods for visually analyzing and representing associations between metabolites and outcomes are borrowed from conventional statistics and other biological fields.[3,4] Creating visualizations that can facilitate the interpretation of multi-level analyses, including information regarding associations among multiple metabolites and multiple outcomes, continues to pose a special challenge. We have therefore developed a visualization technique that expands upon existing approaches to enable the display of results from multiple simultaneous analyses relating metabolites and clinical phenotypes.

## METHODS AND RESULTS

For development of these visualization methods, we used a metabolomics dataset comprising >500 bioactive lipids assayed by high resolution LC-MS in a subset of the Framingham Heart Study offspring cohort (N=1447 participants; see **Supplement** for details).[17] Given the biological importance of lipid metabolites with respect to cardiometabolic disease traits, we used multiple regression models to relate the extended panel of metabolites with clinical traits as part of staged analyses that were conducted in a pre-specified order (**Fig. 1a**). Manhattan plots display P values for each model run, and highlight statistically significant associations of all metabolites with each outcome (**Fig. 1b**). Although colored dots can provide information regarding the directionality of associations, along with corresponding P value, the magnitude of each association is not easily conveyed. This issue can be addressed using parallel heatmaps, a commonly applied visualization approach for high-dimensional 'omics' data, wherein one plot depicts magnitude of effect and directionality (e.g. beta coefficients) while the other depicts corresponding statistical significance (e.g. P value) (**Fig. 1c**). Although such an approach allows for global visualization of data, discerning clear patterns within or between metabolites can be difficult, especially across multiple heatmap plots and with increasing data set size.

To overcome this limitation, we combined the visualization concepts offered by the conventional heatmap and previously reported raindrop plot methods[18] to develop a single technique for visualizing the results of relating multiple metabolites with multiple outcomes. As seen in the adapted 'rain plot' shown in **Fig. 1d**, directionality and magnitude of estimates for the top 50 metabolites associated with the selected outcomes are displayed using a color fill scale and the corresponding significance level is represented by size of the circle (i.e. rain 'droplet'). The metabolites are ordered top to bottom by overall smallest to largest global P value for associations across all clinical traits and outcomes.

The rain plot visualization emphasizes two types of comparisons: (1) between-outcome results, and (2) between-metabolite results. For between-outcome comparisons, visually scanning for vertical patterns of large sized or deeply colored droplets (i.e., droplet 'streams') serves to highlight those outcomes that are the most broadly associated with a given panel of metabolites. As shown in **Fig. 1d**, this particular panel of bioactive lipids appears more globally associated with older age, sex, and Framingham Risk Score. A special feature of the rain plot is its emphasis on potentially important between-metabolite comparisons. For instance, certain metabolites (i.e. 20 and 27) are very strongly associated with sex (**Fig. 1b-d**). For each of these metabolite rows, a visual scan from left to right clarifies the relatively lower degrees of association for these metabolites with other outcomes of interest. The plot also visually clarifies interesting findings between the top-most prioritized metabolites. For instance, Metabolite 1 is positively associated with older age, male sex, and greater metabolic as well as cardiovascular disease risk. Conversely, Metabolite 2 is associated with lower metabolic risk, but is not significantly associated with either age or sex (**Fig. 1d**). The plot also highlights a finding for Metabolite 3 that distinguishes this analyte from Metabolites 1 and 2: while similarly associated with both greater Framingham Risk Score and risk for incident cardiovascular disease events, Metabolite 3 is not associated with prevalent or incident diabetes (**Fig. 1d**). In effect, Metabolite 3 appears associated with both prevalent and future cardiovascular risk through a biological pathway that is likely distinct from diabetes risk. Similar between-metabolite comparisons are possible across the entire plot. The code used to create a rain plot and select parameters for data display is provided for the scientific community (https://github.com/biodatacore/2017.09-rainplots), and this code may be adapted easily for a variety of similar datasets.

## DISCUSSION

As analytical chemistry methods continue to mature, resulting in larger and more complex metabolomics data, there is a growing need for ways to visually understand and interpret the relations of these high-dimensional data with multiple outcomes of interest. Using a dataset of metabolite measures performed in a population scale cohort, we compared several existing visualization techniques (**Supplement**). In this context, we developed and demonstrate the potential utility of a rain plot approach to maximally render the multiple types of information that can be derived from the observed relationships between a panel of metabolites and a set of clinical traits and outcomes. We anticipate that this approach may be further extended and applied to alternate study designs using different types of molecular phenotyping data – as part of the ongoing effort to effectively, efficiently, and feasibly convey the results of large-scale, high-dimensional data analyses.[19,20]

## COMPETING INTERESTS

The authors declare no conflicts of interest.

## CONTRIBUTIONS

Designed the study: MH, TN, JDW, KAL, MJ, and SC. Conducted the experiments: MH, TN, JDW, and KAL. Analysed the data: MH, TN, JDW, KAL, EJD, BVJ, OD, and SC. Provided material, data or analysis tools: JA, BLC, RSV, MGL. Wrote the paper: MH, TN, MJ, SC. All authors read the paper and contributed to its final form.

## FUNDING

This project was supported in part by the National Institutes of Health grants T32-ES007142 (JA), T32-HL007444 (JDW), K01-HL135342 (OD), N01-HC-25195 (RSV, MGL), HHSN268201500001I

(RSV, MGL), R01-HL134168 (SC, MJ), R01-ES027595 (MJ, SC), the American Heart Association CVGPS Pathway Award (MGL, SC, MJ), the Doris Duke Charitable Foundation Grant #2015092 (MJ, SC), the Tobacco Related Disease Research Program (MJ, JDW), and the Frontiers of Innovation Scholars Program (KAL).

## REFERENCES

1 Misra, B. B. & van der Hooft, J. J. Updates in metabolomics tools and resources: 2014-2015. *Electrophoresis* 37, 86-110, doi:10.1002/elps.201500417 (2016).

2 Johnson, C. H., Ivanisevic, J. & Siuzdak, G. Metabolomics: beyond biomarkers and towards mechanisms. *Nat Rev Mol Cell Biol* 17, 451-459, doi:10.1038/nrm.2016.25 (2016).

3 Chia, P. L., Gedye, C., Boutros, P. C., Wheatley-Price, P. & John, T. Current and Evolving Methods to Visualize Biological Data in Cancer Research. *J Natl Cancer Inst* 108, doi:10.1093/jnci/djw031 (2016).

4 Wang, R., Perez-Riverol, Y., Hermjakob, H. & Vizcaino, J. A. Open source libraries and frameworks for biological data visualisation: a guide for developers. *Proteomics* 15, 1356-1374, doi:10.1002/pmic.201400377 (2015).

5 Sugimoto, M. Metabolomic pathway visualization tool outsourcing editing function. *Conf Proc IEEE Eng Med Biol Soc* 2015, 7659-7662, doi:10.1109/EMBC.2015.7320166 (2015).

6 Xia, J., Sinelnikov, I. V., Han, B. & Wishart, D. S. MetaboAnalyst 3.0--making metabolomics more meaningful. *Nucleic Acids Res* 43, W251-257, doi:10.1093/nar/gkv380 (2015).

7 Clasquin, M. F., Melamud, E. & Rabinowitz, J. D. LC-MS data processing with MAVEN: a metabolomic analysis and visualization engine. *Curr Protoc Bioinformatics* Chapter 14, Unit14 11, doi:10.1002/0471250953.bi1411s37 (2012).

8 Grapov, D., Wanichthanarak, K. & Fiehn, O. MetaMapR: pathway independent metabolomic network analysis incorporating unknowns. *Bioinformatics* 31, 2757-2760, doi:10.1093/bioinformatics/btv194 (2015).

9 Matheus, N. *et al.* An easy, convenient cell and tissue extraction protocol for nuclear magnetic resonance metabolomics. *Phytochem Anal* 25, 342-349, doi:10.1002/pca.2498 (2014).

10 Grace, S. C., Embry, S. & Luo, H. Haystack, a web-based tool for metabolomics research. *BMC Bioinformatics* 15 Suppl 11, S12, doi:10.1186/1471-2105-15-S11-S12 (2014).

11 Eichner, J. *et al.* Integrated enrichment analysis and pathway-centered visualization of metabolomics, proteomics, transcriptomics, and genomics data by using the InCroMAP software. *J Chromatogr B Analyt Technol Biomed Life Sci* 966, 77-82, doi:10.1016/j.jchromb.2014.04.030 (2014).

12 Xia, J. & Wishart, D. S. Metabolomic data processing, analysis, and interpretation using MetaboAnalyst. *Curr Protoc Bioinformatics* Chapter 14, Unit 14 10, doi:10.1002/0471250953.bi1410s34 (2011).

13 Chagoyen, M. & Pazos, F. Tools for the functional interpretation of metabolomic experiments. *Brief Bioinform* 14, 737-744, doi:10.1093/bib/bbs055 (2013).

14 Mak, T. D., Laiakis, E. C., Goudarzi, M. & Fornace, A. J., Jr. MetaboLyzer: a novel statistical workflow for analyzing Postprocessed LC-MS metabolomics data. *Anal Chem* 86, 506-513, doi:10.1021/ac402477z (2014).

15 Kuo, T. C., Tian, T. F. & Tseng, Y. J. 3Omics: a web-based systems biology tool for analysis, integration and visualization of human transcriptomic, proteomic and metabolomic data. *BMC Syst Biol* 7, 64, doi:10.1186/1752-0509-7-64 (2013).

16 Karnovsky, A. *et al.* Metscape 2 bioinformatics tool for the analysis and visualization of metabolomics and gene expression data. *Bioinformatics* 28, 373-380, doi:10.1093/bioinformatics/btr661 (2012).

17 Watrous, J. D. *et al.* Visualization, Quantification, and Alignment of Spectral Drift in Population Scale Untargeted Metabolomics Data. *Anal Chem* 89, 1399-1404, doi:10.1021/acs.analchem.6b04337 (2017).

18 *Raindrop Plot*, <http://mc-3.ca/raindrop-plot> (2011).

19 Gehlenborg, N. *et al.* Visualization of omics data for systems biology. *Nat Methods* 7, S56-68, doi:10.1038/nmeth.1436 (2010).

20 O'Donoghue, S. I. *et al.* Visualizing biological data-now and in the future. *Nat Methods* 7, S2-4, doi:10.1038/nmeth.f.301 (2010).

21 Kannel, W. B., Feinleib, M., McNamara, P. M., Garrison, R. J. & Castelli, W. P. An investigation of coronary heart disease in families. The Framingham offspring study. *Am J Epidemiol* 110, 281-290 (1979).

22 Huan, T. & Li, L. Counting missing values in a metabolite-intensity data set for measuring the analytical performance of a metabolomics platform. *Anal Chem* 87, 1306-1313, doi:10.1021/ac5039994 (2015).

23 Grundy, S. M. *et al.* Diagnosis and management of the metabolic syndrome: an American Heart Association/National Heart, Lung, and Blood Institute Scientific Statement. *Circulation* 112, 2735-2752, doi:10.1161/CIRCULATIONAHA.105.169404 (2005).

24 Abraham, T. M., Pencina, K. M., Pencina, M. J. & Fox, C. S. Trends in Diabetes Incidence: The Framingham Heart Study. *Diabetes Care* 38, 482-487, doi:10.2337/dc14-1432 (2015).

25 Wilson, P. W. *et al.* Prediction of coronary heart disease using risk factor categories. *Circulation* 97, 1837-1847 (1998).

26 Niiranen, T. J. *et al.* Prevalence, Correlates, and Prognosis of Healthy Vascular Aging in a Western Community-Dwelling Cohort. *The Framingham Heart Study*, doi:10.1161/hypertensionaha.117.09026 (2017).

27 The R, F. *R: The R Project for Statistical Computing*, <https://www.r-project.org/> (2017).

## FIGURE LEGEND

**Figure 1. Visualization of complex metabolomics data.** For studies that involve multiple staged experiments or statistical models, several options exist for visualizing analysis results. For a set of statistical models (**a**) performed in a large human study, for example, a Manhattan plot (**b**) can display the degree to which a wide panel of metabolites is associated with different outcomes although the magnitude of these associations is not conveyed. Pairing of heatmaps can display magnitude as well as directionality and significance for each metabolite association (**c**), although between-metabolite comparisons of associations across all outcomes is not easily discernible. A 'rain plot' approach (**d**) combines the information from paired heatmaps into a single plot that emphasizes two types of information: (i) between-outcome comparisons, such as the extent to which most metabolites in this panel are associated with certain outcomes (e.g. age, sex) more than others; and, (ii) between-metabolite comparisons, such as the extent to which certain metabolites are associated with an aggregate measure of clinical cardiovascular disease risk (e.g. Framingham risk score) with or without concurrent relations to major component risk factors (e.g. diabetes risk for Metabolites 3 and 4, compared to for Metabolites 1 and 2). Between-metabolite comparisons, in particular, can facilitate identification of potentially important biological differences underlying the observed results of relating multiple metabolites to multiple phenotypes.

**a.**

| Model | Outcome | Covariates in Model |
|---|---|---|
| 1 | age | sex |
| 2 | sex | age |
| 3 | BMI | age, sex |
| 4 | metabolic syndrome | age, sex |
| 5 | prevalent diabetes | age, sex, BMI |
| 6 | incident diabetes | age, sex, BMI, fasting glucose |
| 7 | Framingham risk score | n/a |
| 8 | incident CVD | Framingham risk score |

**b.**

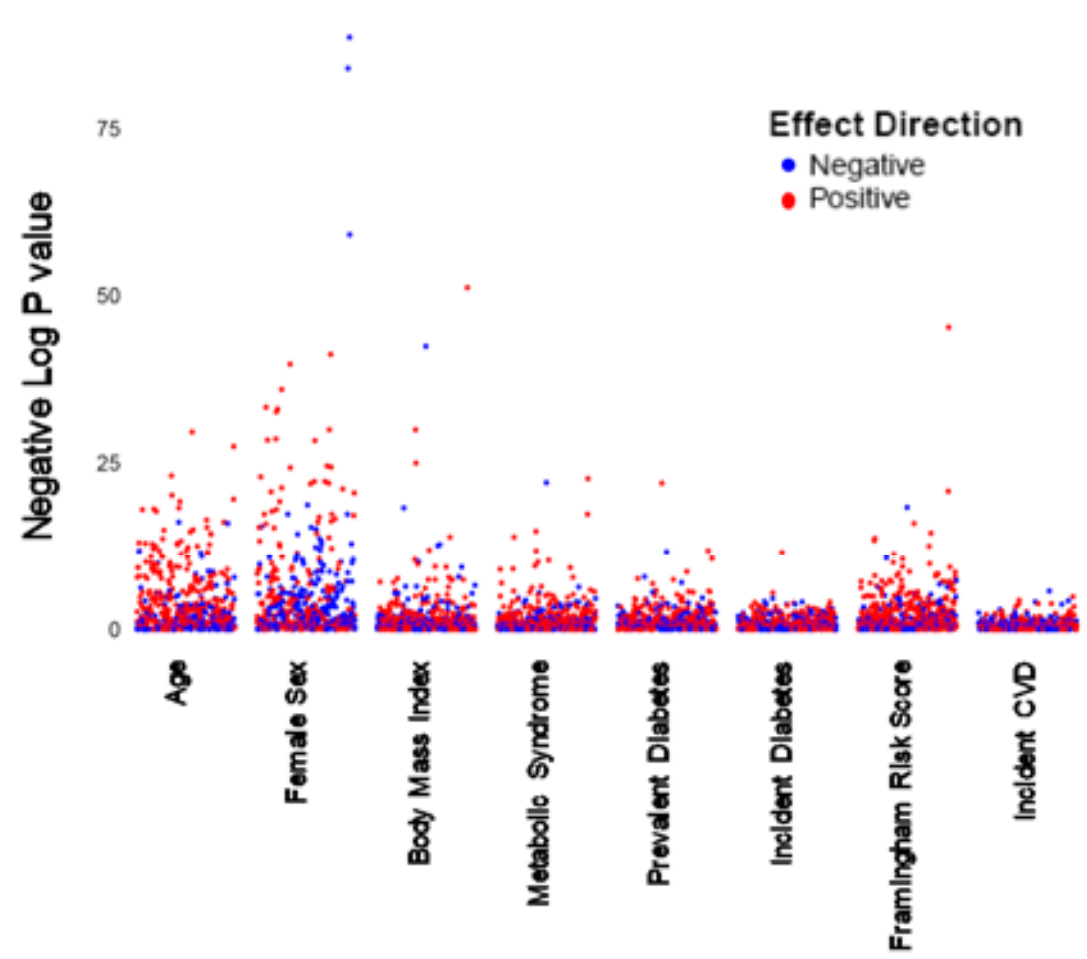

**c.**

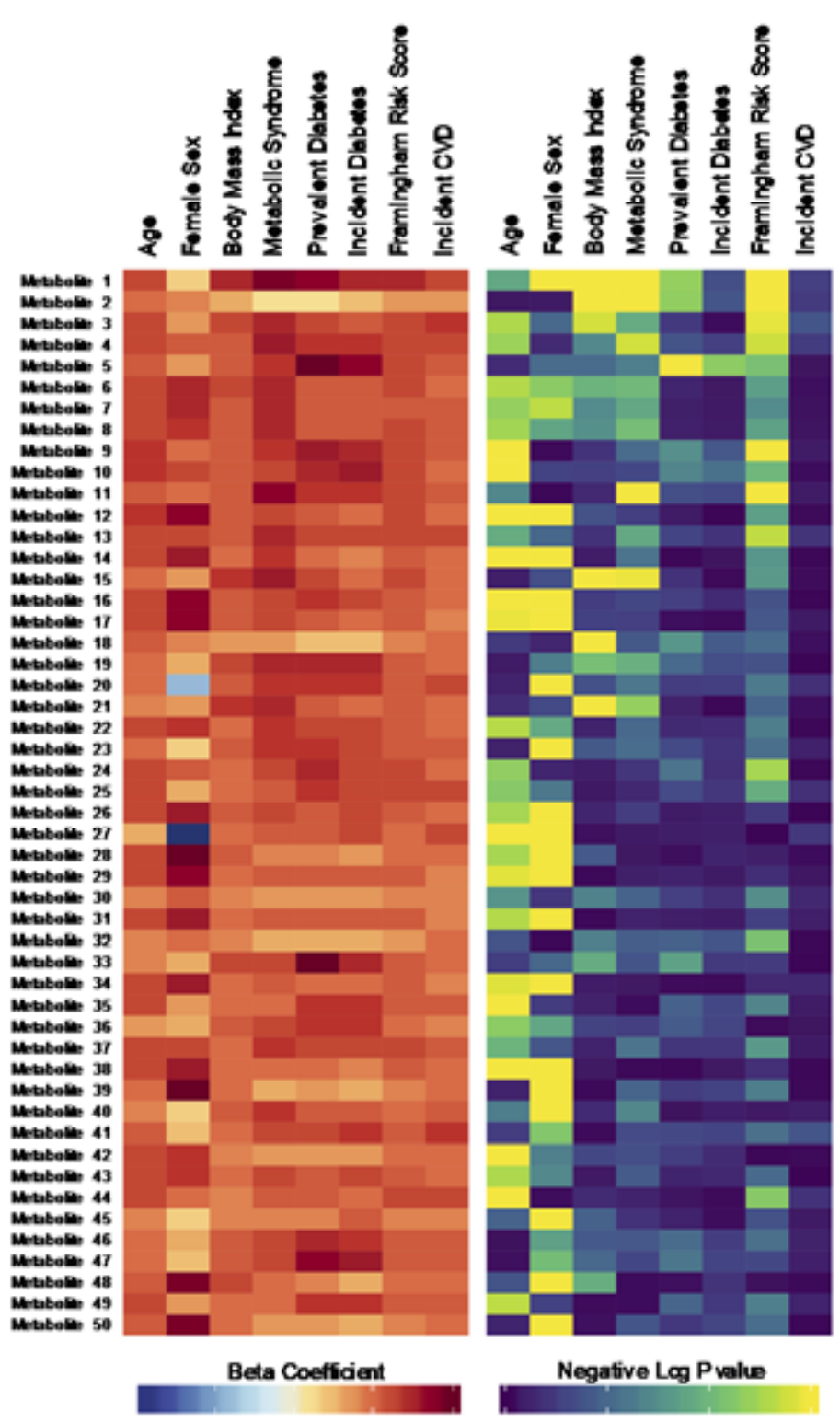

**d.**

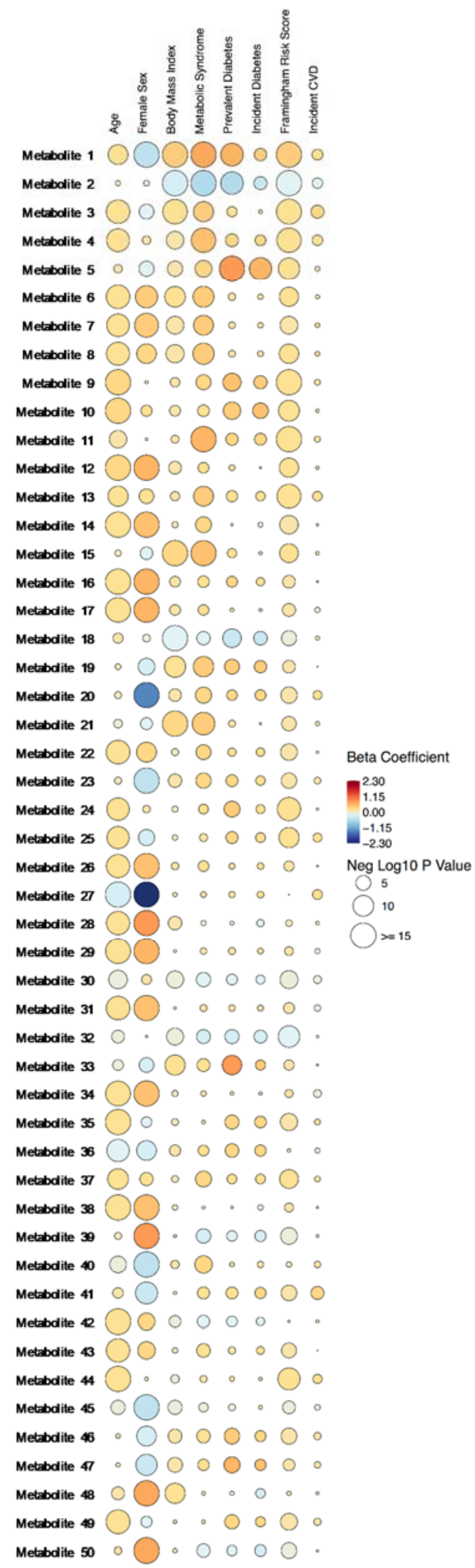

# SUPPLEMENT

## SUPPLEMENTAL METHODS

### Study Cohort and Design

We used high-throughput metabolomics data collected from participants of the Framingham Heart Study (FHS) offspring cohort. In 1971, the Framingham Offspring Study recruited the children and their spouses of the original cohort to participate in a longitudinal epidemiological study.[21] All participants underwent a history, physical, and laboratory assessment at each study examination cycle, approximately every four years. We used a bioactive lipids profiling method with >500 metabolites in N=2895 participants who attended their eighth examination cycle (2005-2008).[17] For this investigation, we analyzed a randomly selected subset comprising N=1447 individuals (mean age 66 [range 40-92] years, 54% women). Informed consent was obtained from all study participants and all study protocols were approved by the Institutional Review Boards of Boston University, Brigham and Women's Hospital, and University of California San Diego.

### Metabolomics Procedures and Data Processing

Nontargeted LC-MS based metabolomics analysis was performed on all available N=2895 plasma samples, according to previously described protocols.[17] In brief, plasma samples were prepared and analyzed using a Thermo Vanquish UPLC coupled to a high resolution Thermo QExactive orbitrap mass spectrometer. Metabolites were isolated from plasma using protein precipitation with organic solvent followed by solid phase extraction. Extracted metabolites were underwent chromatographic separation using reverse phase chromatography whereby samples were loaded onto a Phenomenex Kinetex C18 (1.7um, 2.1x100mm) column and eluted using a 7 minute linear gradient starting with water : acetonitrile : acetic acid (70:30:0.1) and ending with acetonitrile : isopropanol : acetic acid (50:50:0.02). LC was coupled to a high resolution Orbitrap mass analyzer with electrospray ionization operating in negative ion mode, with full scan data acquisition across a mass range of 225 to 650 m/z. Thermo .raw data files were converted to 32-bit centroid .mzXML using Msconvert (Proteowizard software suite), and resulting .mzXML files were analyzed using Mzmine 2.21, as described.[17]

### Statistical Analyses

For the present analyses, 16% of analytes had >10% missing values; we replaced missing values for metabolites with 0.25 x the minimum observed value for that metabolite, as reported previously.[22] Metabolite variables were then natural logarithmically transformed and standardized (mean=0, SD=1) to facilitate cross-metabolite comparisons. We then performed multivariable regression analyses to examine the relation of each metabolite with several clinical traits and outcomes, in the following order: age (adjusted for sex), female sex (adjusted for age), body mass index (adjusted for age and sex), metabolic syndrome defined according to the NCEP III criteria.[23] (adjusted for age and sex), prevalent diabetes as previously defined[24] (adjusted for age and sex), incident diabetes assessed at examination cycle 9 as previously defined [24] (adjusted for age, sex, body mass index, and fasting glucose at examination cycle 8), Framingham Risk Score[25] as a measure of prevalent cardiovascular risk assessed at examination cycle 8, and incident hard cardiovascular disease [CVD] as defined previously[26] (adjusted for Framingham Risk Score at exam 8).

### Data Visualization

We displayed the results of these relational analyses using a variety of techniques, including Manhattan plots (for one outcome at a time, with results ordered by mass-to-charge [m/z] value), bar and scatter plots (for one outcome at a time, with bars representing magnitude and directionality of estimates, and scatter dots representing P values), heatmaps (p’s or beta’s only), rain plot (beta's, p's, trends across a panel). We created heatmaps and rainplots with metabolites both unclustered and clustered based on hierarchical clustering (R hclust function). We combined the visualization concepts captured by the conventional heatmap and previously reported raindrop plot[18] methods to develop a rainplot approach to visualizing metabolomics association analyses results. Details regarding the coding schema used to develop the rainplot approach and select specific parameters for the type of data displayed, for a given set of outcomes, are provided at: https://github.com/biodatacore/2017.09-rainplots. All analyses and data visualizations were performed using R v3.4.1and RStudio v1.0.153, and visualizations were developed using ggplot2.[27]

## SUPPLEMENTAL RESULTS

The study cohort comprised N=1447 individuals (age 66±9 years, 54% women), including 54% women who were aged 66±9 years with average body mass index 28±5 kg/m2, Framingham Risk Score 9±4 points, 62% metabolic syndrome, and 14% diabetes. All approaches to visualizing the results of association analyses demonstrated a range in the degree to which different metabolites were related to the different outcomes of interest. The extent and type of information conveyed varied across the visualization techniques, as summarized in **Table S1** and detailed below.

The Manhattan plots display P values for each model run, and highlight statistically significant associations for all outcomes across all the metabolites analyzed and ordered from left to right along the x axis by mass-to-charge ratio (m/z). In the individual plots displayed, select metabolites associations appear very significant for almost all outcomes (**Figure S1**). The marked significance of metabolite associations with sex as well as BMI, compared with the other outcomes, is more clearly demonstrated when the results for all outcomes are displayed in a faceted plot with a shared y axis. The directionality of associations is not conveyed, however, such that it is unclear from these plots if the most significant sex associations represent metabolites with elevated circulating levels in men versus women.

The bar and scatter plots display both the directionality and magnitude of associations, for each outcome, along with P values for the top 50 associated metabolites rank ordered by significance from left to right (**Figure S2**). These plots are rendered separately for each outcome, with variable x and y axis limits, although these could be aligned to match within a faceted plot displaying results for multiple outcomes. The extent to which the same metabolite is positively or negatively associated with different outcomes is not as easily discernible. However, the overall trend of generally more positive or more negative associations observed between a given outcome (e.g., age and Framingham Risk Score) and a large panel of metabolites is clearly displayed. Exceptions to such trends are also highlighted. For instance, although most of the more significant sex-related metabolites are associated with being female (coded as 2) versus male (coded as 1), the top sex-related metabolites are associated with being male and the magnitude of these estimates is large.

There are several ways to use the heatmap technique to visualize metabolite-phenotype associations, with and without clustering. A common approach is to display directionality and magnitude of estimates for different outcomes on a color scale, with or without magnitude of corresponding P values in a separate parallel plot on a different color scale (**Figure S3**). While the ordering of clinical phenotypes may be pre-specified, the ordering of metabolites may be clustered (Figure 4A) or unclustered, and potentially ordered by smallest to largest P value for top hits (Figure 4B), m/z ratio (Figure 4C), or by biological pathways if these are known.

As an extension from the heatmap approach to visualizing results for multiple outcomes in relation to multiple metabolites, we also developed a rain drop visualization method. As shown in **Figure S4**, directionality and magnitude of estimates for the top 50 metabolites associated with the selected outcomes are displayed by color fill scale (red for positive directionality, blue for negative directionality, and color shade for effect size) with corresponding P values represented by size of the circle (or rain “droplet”). The metabolites are ordered top to bottom by overall smallest to larges P value for associations across all clinical traits and outcomes. The figure shows vertical patterns featuring more prominent droplet “streams” for certain outcomes (e.g. age, BMI, and Framingham Risk Score), that highlight the degree to which this panel of metabolites is generally positively associated with older age and greater BMI. The figure also shows horizontal patterns of interest, which are potentially important for comparing the extent to which individual metabolites (profiled as part of the same MS method or panel of metabolites) are or are not related in the same way with a set of clinical traits. For instance, the first row metabolite is positively associated with older age, male sex, and greater metabolic as well as cardiovascular disease risk. By contrast, the second row metabolite is associated with lower metabolic risk in the absence of any significant relationship with either age or sex. Furthermore, the rain plot highlights a finding for the third row metabolite that distinguishes this analyte from the first and second row metabolite. This third row metabolite is similarly associated with both greater Framingham Risk Score and risk for incident cardiovascular disease events; however, unlike for the first and second row metabolites, this third row metabolite is not associated with prevalent or incident diabetes. Thus, the rain plot visualization clarifies the analytical results suggesting that the third row metabolite is associated with both prevalent and future cardiovascular risk via a biological pathway that is much less likely to involve diabetes, when compared to pathways represented by other metabolite associations. These types of distinguishing characteristics, seen on a metabolite by metabolite basis, are not as easily discernible from the other visualizations.

## SUPPLEMENTAL REFERENCES

1 Kannel, W. B., Feinleib, M., McNamara, P. M., Garrison, R. J. & Castelli, W. P. An investigation of coronary heart disease in families. The Framingham offspring study. *Am J Epidemiol* 110, 281-290 (1979).

2 Watrous, J. D. *et al.* Visualization, Quantification, and Alignment of Spectral Drift in Population Scale Untargeted Metabolomics Data. *Anal Chem* 89, 1399-1404, doi:10.1021/acs.analchem.6b04337 (2017).

3 Huan, T. & Li, L. Counting missing values in a metabolite-intensity data set for measuring the analytical performance of a metabolomics platform. *Anal Chem* 87, 1306-1313, doi:10.1021/ac5039994 (2015).

4 Grundy, S. M. *et al.* Diagnosis and management of the metabolic syndrome: an American Heart Association/National Heart, Lung, and Blood Institute Scientific Statement. *Circulation* 112, 2735-2752, doi:10.1161/CIRCULATIONAHA.105.169404 (2005).

5 Abraham, T. M., Pencina, K. M., Pencina, M. J. & Fox, C. S. Trends in Diabetes Incidence: The Framingham Heart Study. *Diabetes Care* 38, 482-487, doi:10.2337/dc14-1432 (2015).

6 Wilson, P. W. *et al.* Prediction of coronary heart disease using risk factor categories. *Circulation* 97, 1837-1847 (1998).

7 Niiranen, T. J. *et al.* Prevalence, Correlates, and Prognosis of Healthy Vascular Aging in a Western Community-Dwelling Cohort. *The Framingham Heart Study*, doi:10.1161/hypertensionaha.117.09026 (2017).

8 *Raindrop Plot*, <http://mc-3.ca/raindrop-plot> (2011).

9 The R, F. *R: The R Project for Statistical Computing*, <https://www.r-project.org/> (2017).

## SUPPLEMENTAL FIGURE LEGENDS

**Figure S1. Manhattan plots.** Panel A shows Manhattan plots displaying the distribution of P values for associations between metabolites (ordered by mass-to-charge ratio along the x-axis) and each clinical phenotype separately. Panel B shows Manhattan plots for all clinical traits combined into a single visualization with a single y-axis displaying a common maximal value to facilitate comparisons between results across outcomes.

**Figure S2. Bar and scatter plots.** Bar and scatter plots display both the beta coefficients and corresponding P values for associations between metabolites and different clinical traits or outcomes (one per plot) with the order of metabolites ordered left to right from smallest to largest P value. Panel A displays results for all analyses, and Panel B displays results for the top 50 associated metabolites.

**Figure S3. Parallel heat maps.** Parallel heat maps display beta coefficients (left plot) and corresponding P values (right plot) for the top 50 metabolites associated with the clinical traits and outcomes listed as column headings. Results are ordered top to bottom by smallest to largest P value (Panel A), mass-to-charge ratio (Panel B), and by clustering (Panel C).

**Figure S4. Rain plots.** Rain plots display the directionality of beta coefficients by color (red for positive, blue for negative), the magnitude of effect by color scale (deeper color for larger magnitude of effect), the significance of association by circle size (larger circle size for greater statistical significance and smaller P value), and the degree to which one or more clinical outcomes is associated with multiple metabolites measured using the same method. Results are ordered top to bottom by smallest to largest P value (Panel A), mass-to-charge ratio (Panel B), and by clustering (Panel C).

**Table S1.** Dimension of information offered by different visualization methods.

| Visualization method | Significance of associations with an outcome | Magnitude and directionality of associations with an outcome | Clustering | Significance of associations with multiple outcomes | Magnitude and directionality of associations with multiple outcomes |
|---|---|---|---|---|---|
| **Manhattan plot** | X | | | X | |
| **Bar and scatter plots** | X | X | | | |
| **Heatmap** | | X | X | X | |
| **Rain plot** | X | X | X | X | X |

**Figure S1A.**

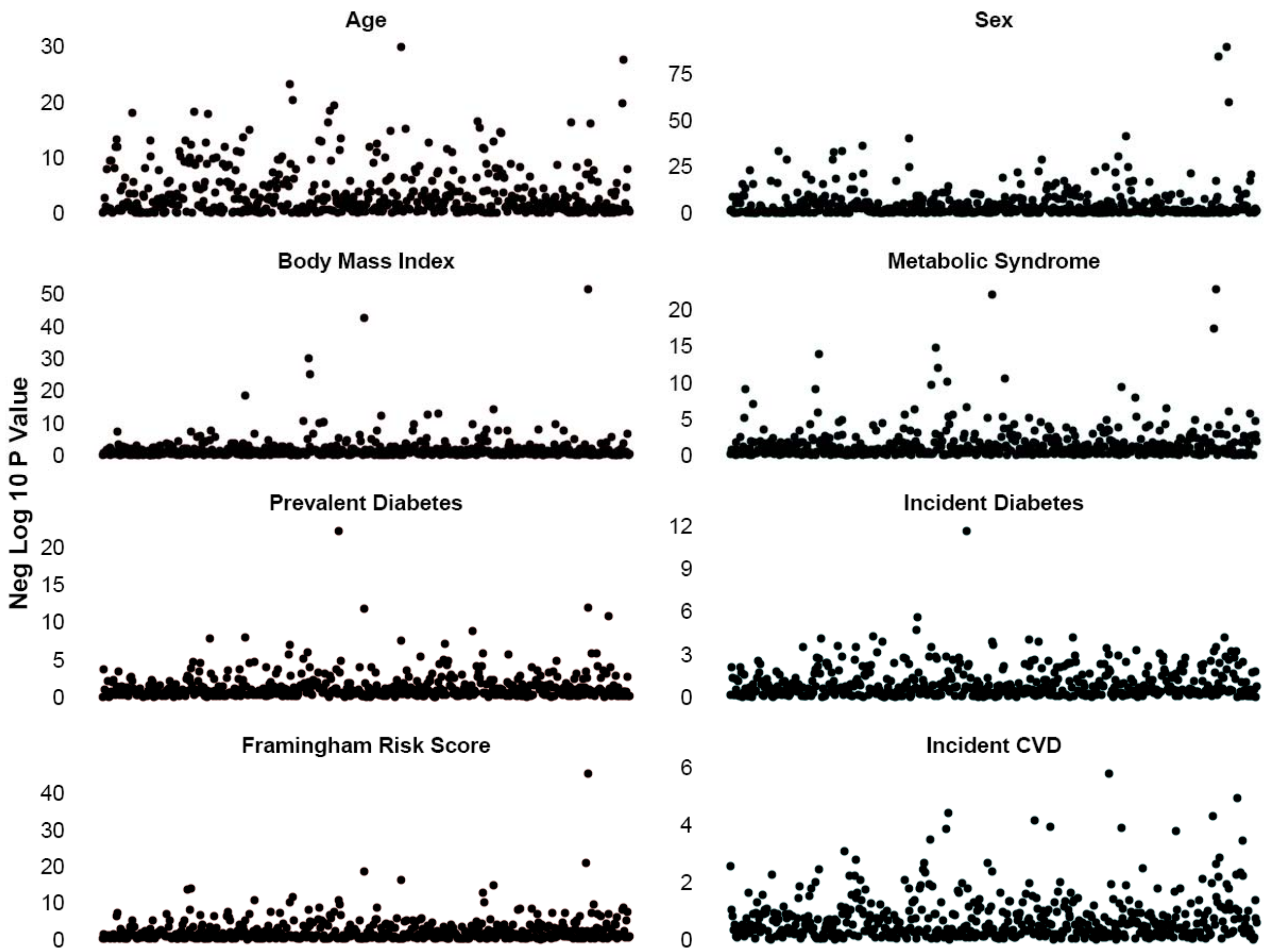
Age
Sex
Body Mass Index
Metabolic Syndrome
Prevalent Diabetes
Incident Diabetes
Framingham Risk Score
Incident CVD
Neg Log 10 P Value
30
20
10
0
75
50
25
0
50
40
30
20
10
0
20
15
10
5
0
20
15
10
5
0
12
9
6
3
0
40
30
20
10
0
6
4
2
0

**Figure S1B.**

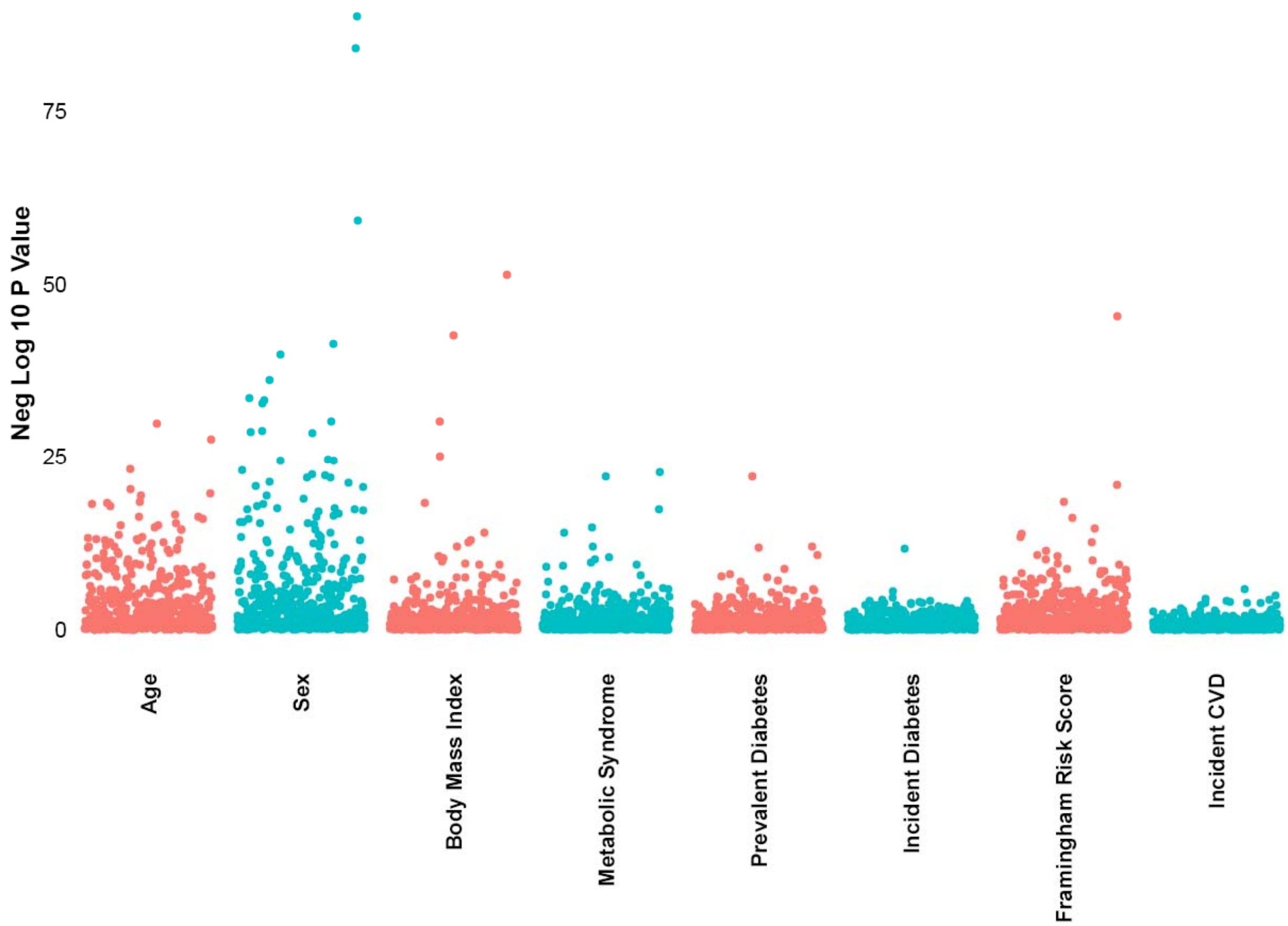

Neg Log 10 P Value
75
50
25
0
Age
Sex
Body Mass Index
Metabolic Syndrome
Prevalent Diabetes
Incident Diabetes
Framingham Risk Score
Incident CVD

**Figure S2A.**

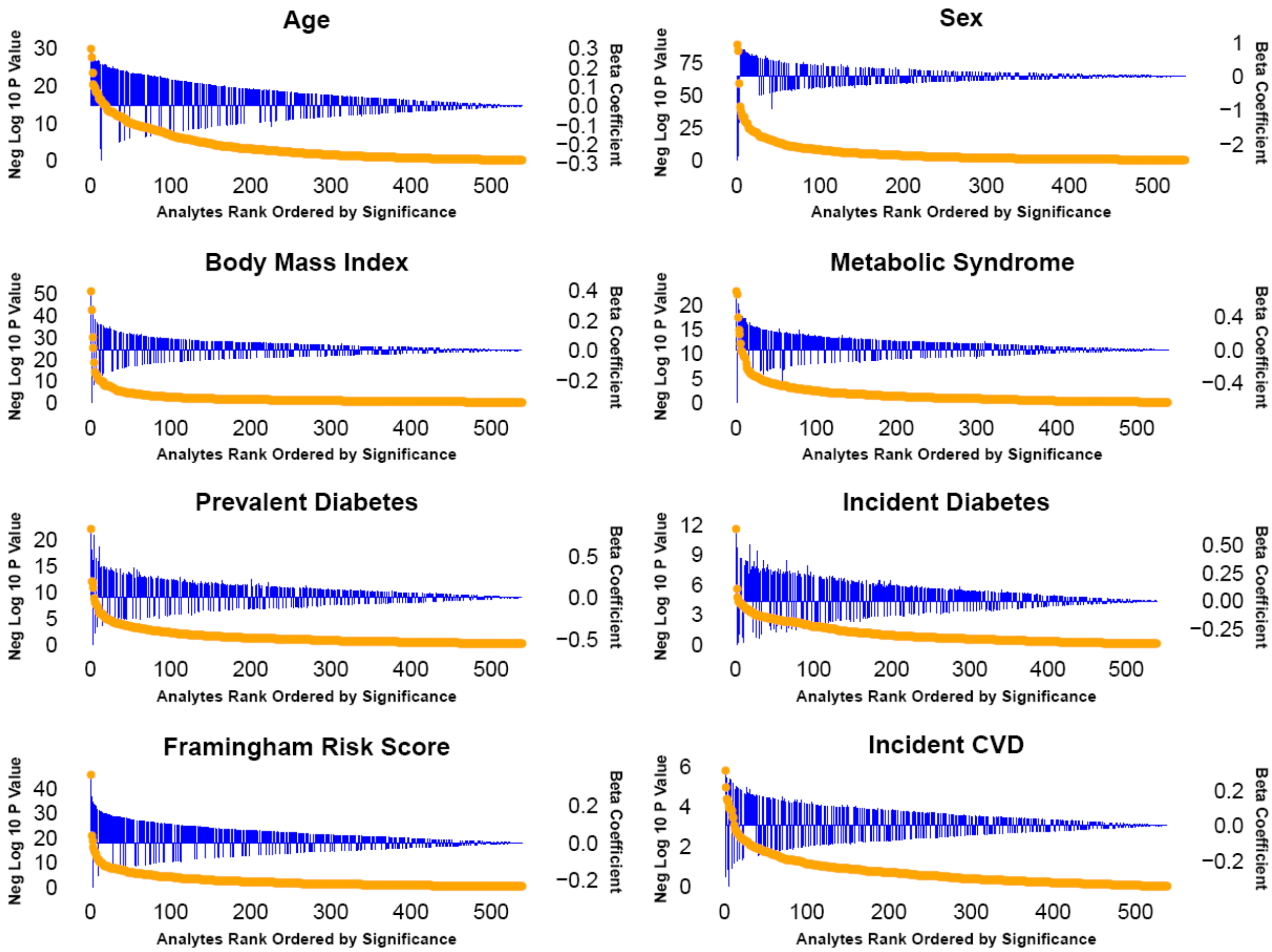

Age
Sex
Body Mass Index
Metabolic Syndrome
Prevalent Diabetes
Incident Diabetes
Framingham Risk Score
Incident CVD
Neg Log 10 P Value
Beta Coefficient
Analytes Rank Ordered by Significance

**Figure S2B.**

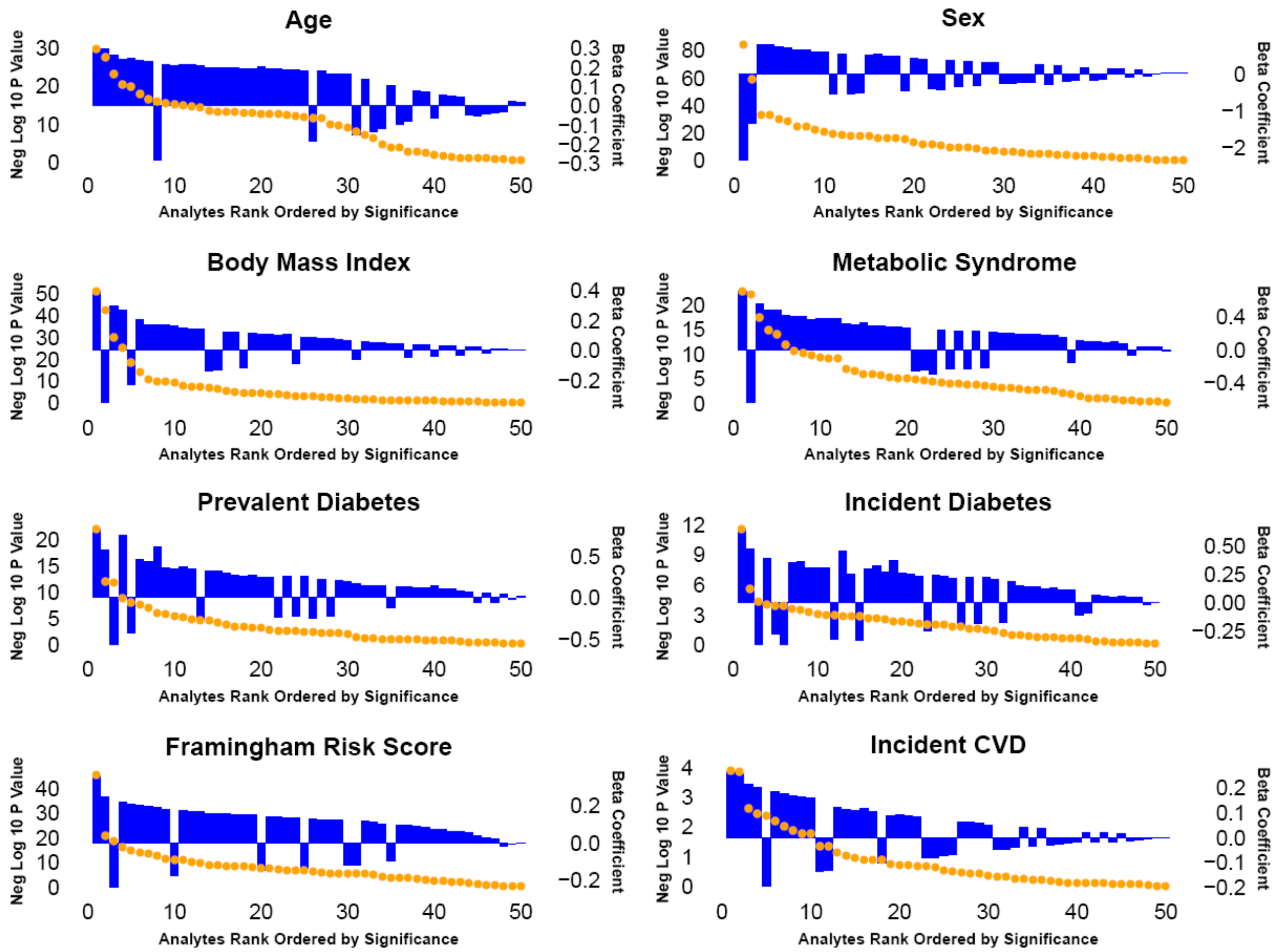
Age
Sex
Body Mass Index
Metabolic Syndrome
Prevalent Diabetes
Incident Diabetes
Framingham Risk Score
Incident CVD
Neg Log 10 P Value
Beta Coefficient
Analytes Rank Ordered by Significance

**Figure S3A.**

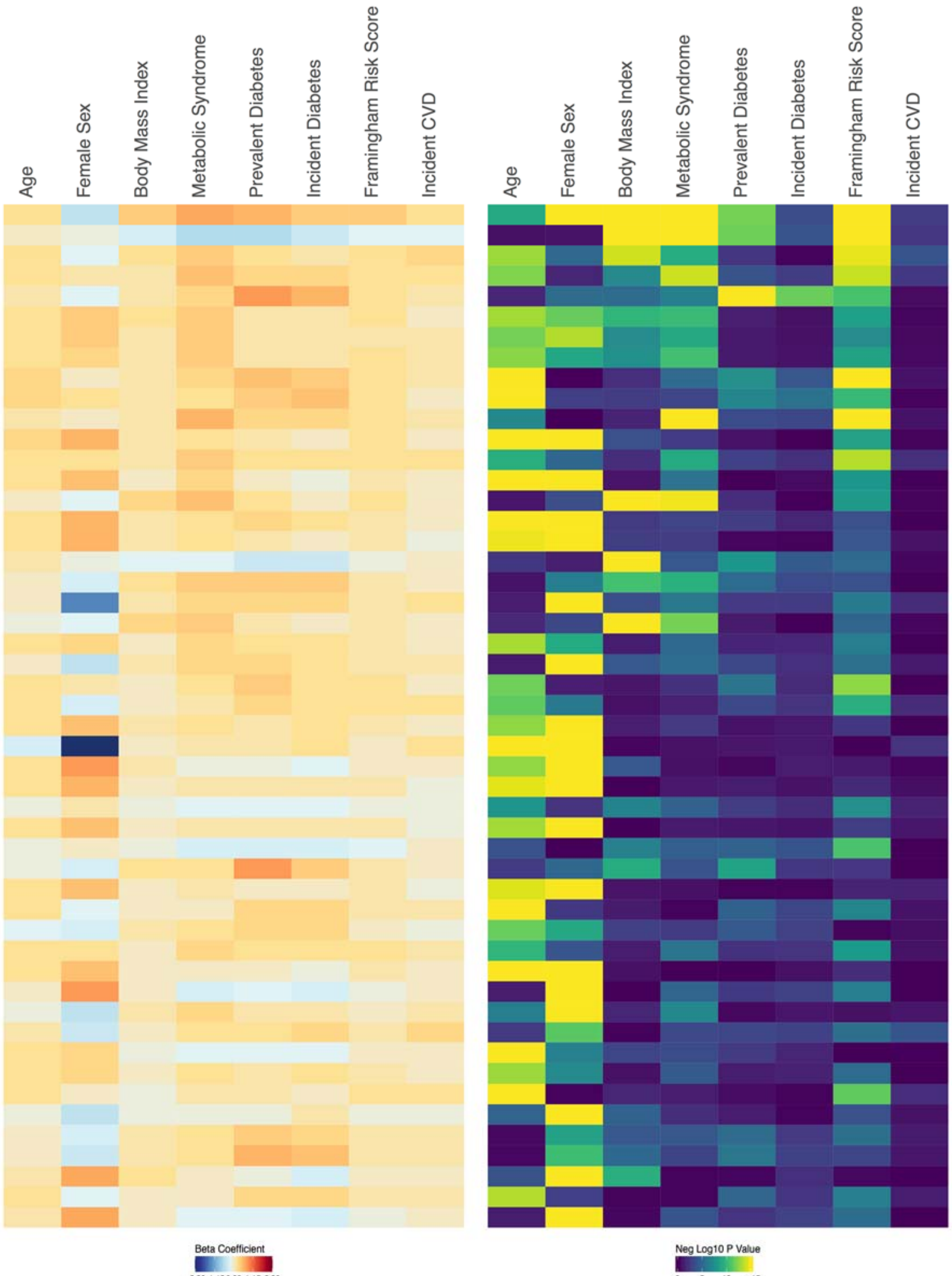
Age
Female Sex
Body Mass Index
Metabolic Syndrome
Prevalent Diabetes
Incident Diabetes
Framingham Risk Score
Incident CVD
Age
Female Sex
Body Mass Index
Metabolic Syndrome
Prevalent Diabetes
Incident Diabetes
Framingham Risk Score
Incident CVD
Beta Coefficient
−2.30 −1.15 0.00 1.15 2.30
Neg Log10 P Value
0 5 10 >=15

**Figure S3B.**

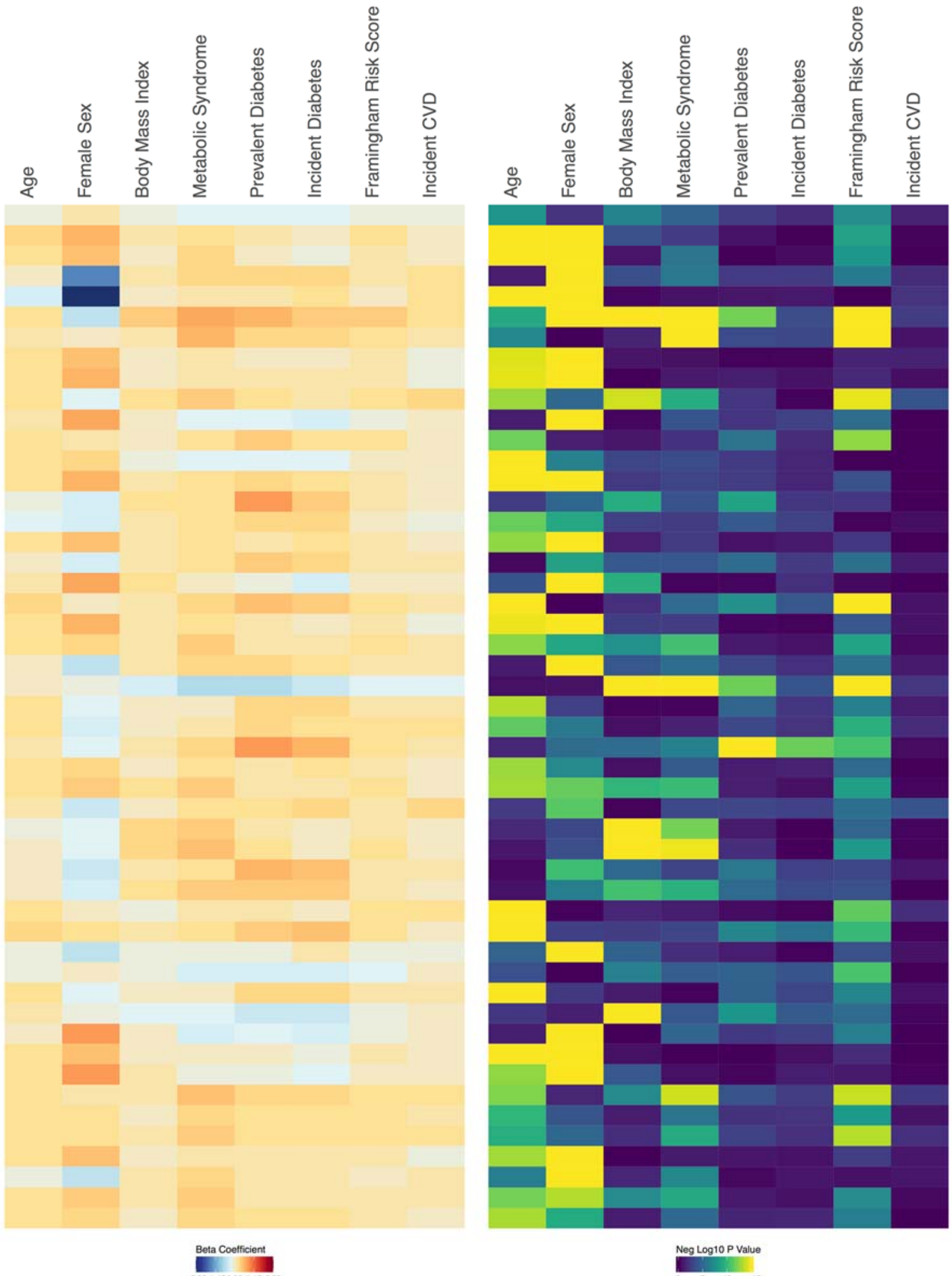
Age
Female Sex
Body Mass Index
Metabolic Syndrome
Prevalent Diabetes
Incident Diabetes
Framingham Risk Score
Incident CVD
Age
Female Sex
Body Mass Index
Metabolic Syndrome
Prevalent Diabetes
Incident Diabetes
Framingham Risk Score
Incident CVD
Beta Coefficient
−2.30 −1.15 0.00 1.15 2.30
Neg Log10 P Value
0 5 10 >=15

**Figure S3C.**

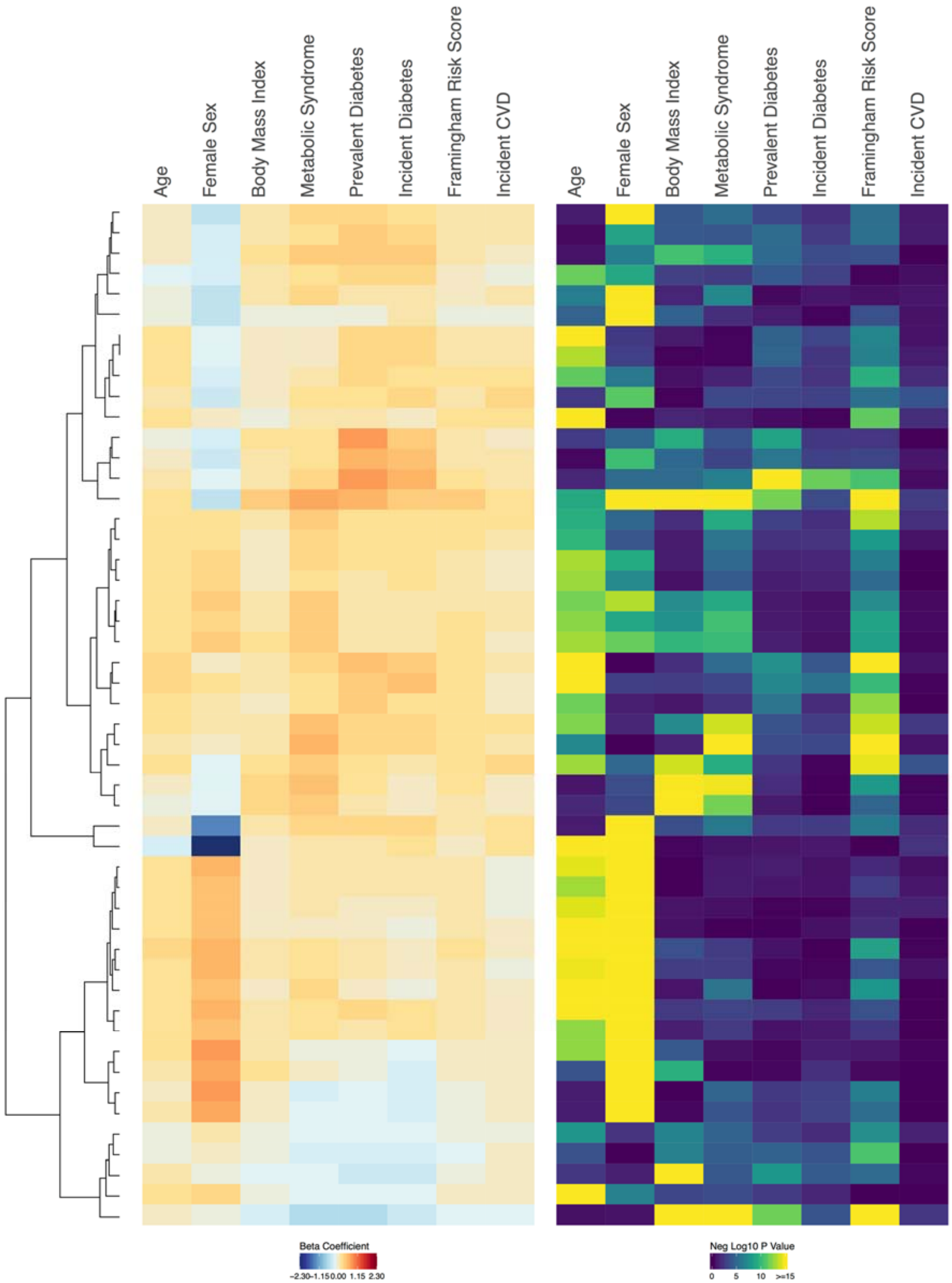
Age
Female Sex
Body Mass Index
Metabolic Syndrome
Prevalent Diabetes
Incident Diabetes
Framingham Risk Score
Incident CVD
Age
Female Sex
Body Mass Index
Metabolic Syndrome
Prevalent Diabetes
Incident Diabetes
Framingham Risk Score
Incident CVD
Beta Coefficient
−2.30 −1.15 0.00 1.15 2.30
Neg Log10 P Value
0 5 10 >=15

**Figure S4A.**

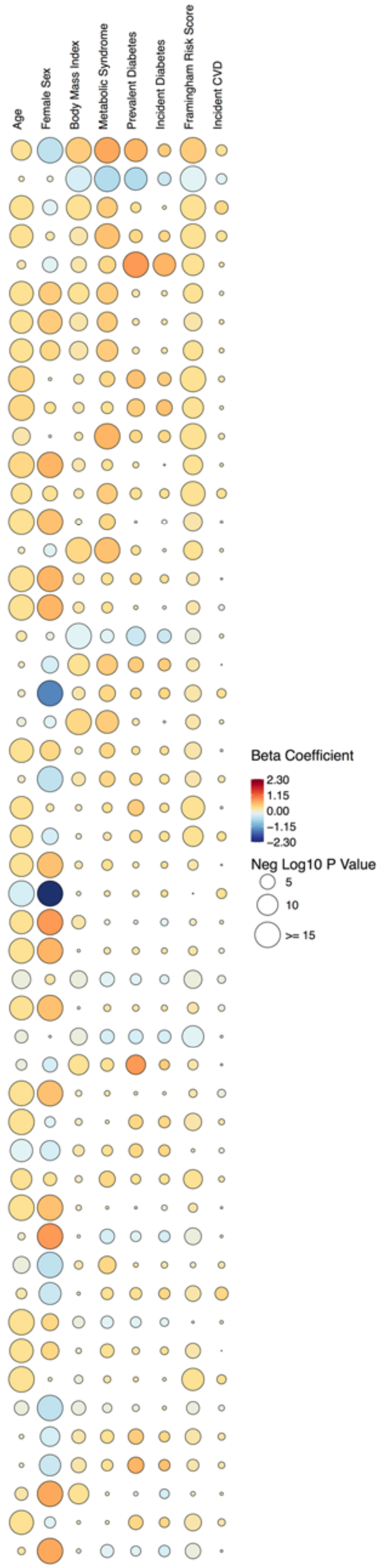

Age
Female Sex
Body Mass Index
Metabolic Syndrome
Prevalent Diabetes
Incident Diabetes
Framingham Risk Score
Incident CVD
Beta Coefficient
2.30
1.15
0.00
−1.15
−2.30
Neg Log10 P Value
5
10
>= 15

**Figure S4B.**

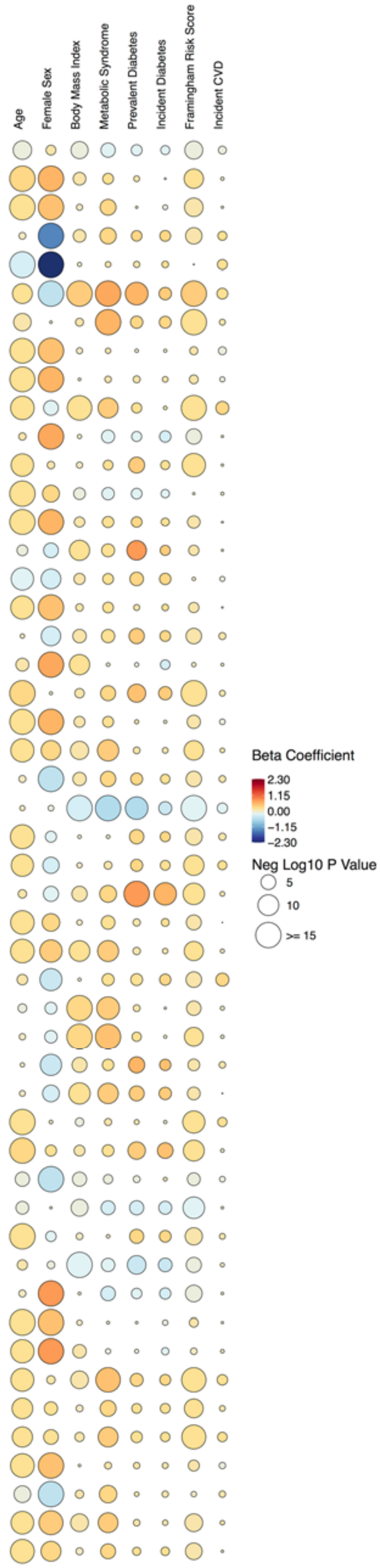

Age
Female Sex
Body Mass Index
Metabolic Syndrome
Prevalent Diabetes
Incident Diabetes
Framingham Risk Score
Incident CVD
Beta Coefficient
2.30
1.15
0.00
−1.15
−2.30
Neg Log10 P Value
5
10
>= 15

**Figure S4C.**

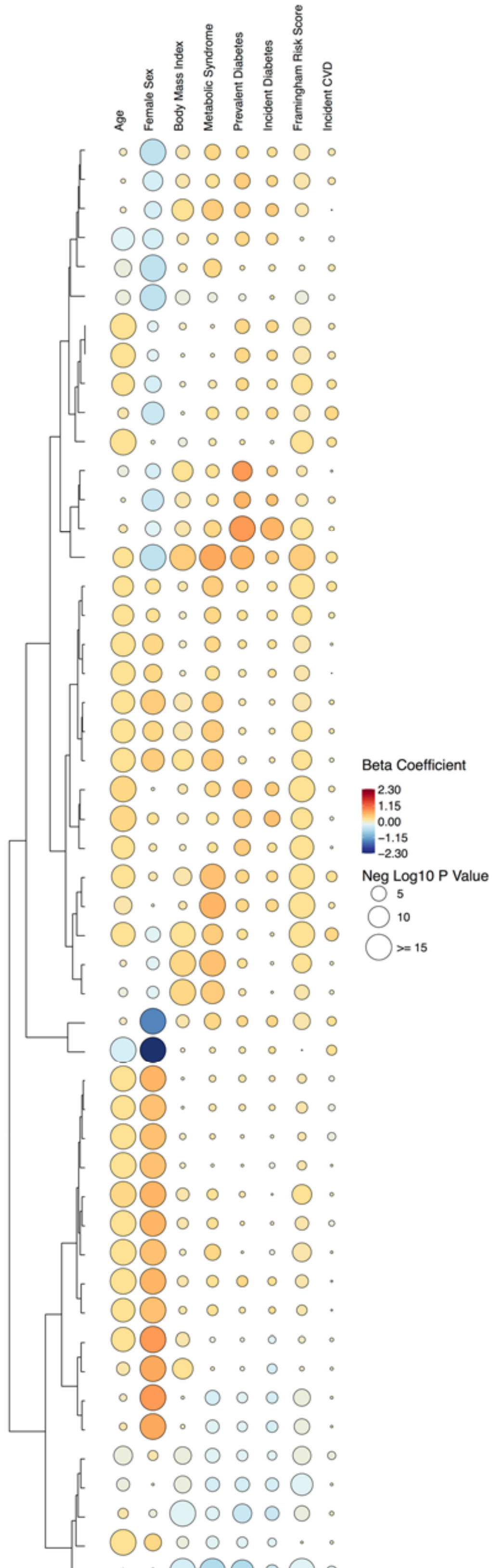
Age
Female Sex
Body Mass Index
Metabolic Syndrome
Prevalent Diabetes
Incident Diabetes
Framingham Risk Score
Incident CVD
Beta Coefficient
2.30
1.15
0.00
−1.15
−2.30
Neg Log10 P Value
5
10
>= 15